\documentclass[%
 reprint,
superscriptaddress,
 showpacs,preprintnumbers,
 bibnotes,
 amsmath,amssymb,
 aps,
 prb,
]{revtex4-1}

\usepackage{graphicx}
\usepackage{dcolumn}
\usepackage{bm}
\usepackage{epstopdf}

\usepackage{color}
\usepackage{float}
\usepackage{amsfonts}

\begin{document}


\title{Electron spin rotations induced by oscillating Rashba interaction in a quantum wire}
\author{J. Paw\l{}owski}
\affiliation{
	Faculty of Physics and Applied Computer Science,
	AGH University of Science and Technology, Krak\'{o}w, Poland}
\author{P. Szumniak}
\affiliation{Department of Physics, University of Basel, Klingelbergstrasse 82, 4056 Basel, Switzerland}
\affiliation{
	Faculty of Physics and Applied Computer Science,
	AGH University of Science and Technology, Krak\'{o}w, Poland}
\author{S. Bednarek}
\affiliation{
	Faculty of Physics and Applied Computer Science,
	AGH University of Science and Technology, Krak\'{o}w, Poland}

\date{\today}

\begin{abstract}
A novel method and nanodevice are introduced that allows to rotate the single electron spin confined in a gated electrostatic InSb nanowire quantum dot. 
Proposed method does not require application of any (oscillating or static) external magnetic fields. 
Our proposal instead employs spatial and time modulation of confining potential induced by electric gates, which, in turn leads to oscillating Rashba type spin-orbit coupling. 
Moving electron back and forth in such a variable Rashba field allows for realization of spin rotations around two different axes separately without using an external magnetic field.
The results are supported by realistic three-dimensional time dependent Poisson-Schr\"{o}dinger calculations for systems and material parameters corresponding to experimentally accessible structures.

\vspace{3 mm}
\end{abstract}

\pacs{71.70.Ej, 73.21.La, 03.67.Lx}

\maketitle


\section{INTRODUCTION}
There is currently great interest in studying spin related phenomena in semiconductor quantum nanowires with spin-orbit interaction (SOI). On the one hand there is a novel fundamental physics at the nanoscale and on the other hand one expects applications in terms of spintronics \cite{s1, s5a,zutic,8} and spin-based quantum information processing \cite{CKRevl}. Furthermore nanowires with SOI are considered to be a promising platform for realizing Majorana fermions in condensed matter\cite{alicea2012new, PhysRevB.79.094504, PhysRevLett.105.077001, PhysRevLett.105.177002, mourik2012signatures, deng2012anomalous, das2012zero, rokh2012gacje, PhysRevB.83.094525, PhysRevLett.109.236801, PhysRevB.87.024515} and, very recently, also for parafermions \cite{PhysRevB.89.115402, PhysRevB.90.045118}, which can be exploited for realization of topological quantum computing \cite{alicea2012new, RevModPhys.80.1083}.

Physical realization of quantum computers requires fulfillment of a number of challenging criteria\cite{s3}. A fragile quantum state has to be coherent for sufficient long time which usually requires its isolation from the environment. On the other hand it has to be externally manipulated. For these purposes, the electron spin in semiconductor quantum dots was suggested as a promising candidate\cite{s4}. There are a number of experiments in which the electron spin confined in quantum dots  is initialized, manipulated, stored, and read out\cite{s7,s8,s9,s10,s11,s12,s13,s14,s15,1,2,13}.

{Furthermore, the ability to address individual solid state spin qubits in a quantum register is essential for realization of scalable quantum computer architectures.  The spin manipulation techniques based on interplay of SOI and oscillating electrical fields are very promising for achieving this goal. These are so-called electric dipole spin resonance (EDSR) techniques proposed in Refs.~[\onlinecite{ed1,ed2,ed3}]. Electric fields can be very precisely and locally applied by means of nanometer size metallic gates. This is in contrast to oscillating magnetic fields, which are difficult to confine in nanoscale.}

The EDSR driven spin rotations were demonstrated in the state of the art experiments for electrons trapped in the planar electrostatic GaAs quantum dots\cite{12} and for InAs or InSb nanowire quantum dots defined by local gating\cite{13,18}. In the latter case much shorter operation times were obtained. This is because the typical SOI couplings in InAs and InSb nanowires are much larger than those in GaAs quantum dots defined electrostatically in planar heterostructures. Recent experimental results show that by properly designing electrodes, the Rashba SOI in InSb nanowire can reach a huge value\cite{14} corresponding to a spin-orbit energy of $0.25$-$1$~meV. However one has to remember that reliable and precise estimate of SOI is a very challenging task\cite{diego}.

In the EDSR technique one force a cyclic motion of the electron back and forth.  Due to the fact that at zero external magnetic field the system has a time-reversal symmetry (spin-orbit interaction does not lift this symmetry), the rotation of the spin associated with the motion in one direction is opposite to rotation for electron moving back. As a result, in 1D system one does not get effective spin rotation for electron which returns to the initial position on the same path. To lift this symmetry, one have to apply external magnetic field.

Hoverer, fully magnetic free spin rotations can be realized for the electrons confined in the 2D electrostatic quantum dots exploiting single charge motion along the two-dimensional closed curve and SOI. 
Then the electron deflected from the equilibrium position moves back (on different trajectory) to initial position getting different spin state, in a zero magnetic field.
Several semiconductor nanodevices capable to realize such a magnetic free operation has been proposed, where spin rotation can be realized by transporting electron or hole along closed-loop once \cite{15, HH1, HH2} for induced quantum wires, or repeatedly\cite{16}, for the electrostatic quantum dots of currently achievable designs and sizes\cite{17}. 
{The particularly promising experimental realization of fully magnetic-free control of electron spin was demonstrated very recently\cite{sanada}. The single electron spin rotations were realized by transporting an electron by a surface acoustic waves in the presence of spin-orbit interaction in 2D micrometer size system, however not in the nanoscale.}

The question now emerges: is it possible to realize effective spin rotations in a 1D nanowire, without the use of an external magnetic fields?
As mentioned earlier, in order to obtain effective spin rotation, the electron deflected from the initial position has to go back on different trajectory. In a nanowire, where in the transverse degrees of freedom are frozen, such a motion is impossible.
Within the paper we show that by using the SOI of Rashba type which can be modulated by an alternating electric field one can realize electron spin rotations without magnetic field. It turns out that if, during the oscillatory motion of the electron in the wire, the confinement potential is appropriately modulated in the transverse direction, the Rashba SOI coupling will be different for electron moving in one direction than for electron moving back. Here we demonstrate that such an oscillatory change of the spin-orbit coupling allows for spin rotations for electron moving back and forth along the nanowire.

Effects of the modulated Rashba SOI have recently been studied in different structures in the context of spin current polarization due to spatially non-uniform Rashba field in heterostructures\cite{20}, and quantum wires\cite{26}. Furthermore oscillating Rashba coupling has been exploited to propose the fast spin control in a single-\cite{22} and two-electron\cite{23} quantum dots as well as for two-electron spin-density separation \cite{24}. Such spins separation may be used for a single spin initialization in a zero magnetic field\cite{24}.
Effects of time dependent RSO coupling on the spin polarized currents has been studied for graphene monolayers \cite{21} and for the two-dimensional superlattices\cite{25}. Nonhomogeneous Rashba interaction also leads to the possibility of controlling the magnetic anisotropy of thin ferromagnetic layers using an electric field\cite{7}.
Moreover, interesting effects of non uniform Rashba SOI on Majorana bound states in hybrid semiconductor Rashba nanowires has been investigated in Ref.~[\onlinecite{Jelena}].

In the recent papers\cite{slo1,slo2}, the authors have shown the possibility of spin-qubit manipulations without using magnetic field in the presence of time-dependent SOI. The calculations were carried out analytically for one-dimensional systems and exact analytical solutions were obtained. 

{Our study is more related to the experimental structures since we include realistic 3D structure of the nanowire with material interfaces and the geometry and position of the device electrodes. This is all very important for detail description of the realistic nanoscale system and was not considered in the Refs.~[\onlinecite{slo1,slo2}]. Furthermore, in the present work, we consider the nanodevices similar to those from the experimental set-ups~\cite{1,2}, which are based on InSb nanowires with a diameter of 50~nm. According to our calculations the 1D approximation is valid only for nanowires with a diameter below 10 nm but not for such thick wires. Thus in order to describe electron spin physics of proposed nanodevices correctly we perform precise three-dimensional self-consistent time dependent Poisson-Schr\"{o}dinger calculations. 
Moreover presented 3D modeling can correctly describe essential for our method changes of the local magnitude of the Rashba SOI, extracted from the electric field present in the system.}



\section{DEVICE AND CALCULATION METHOD}

\begin{figure}[tbh!]
\includegraphics[width=7.5cm]{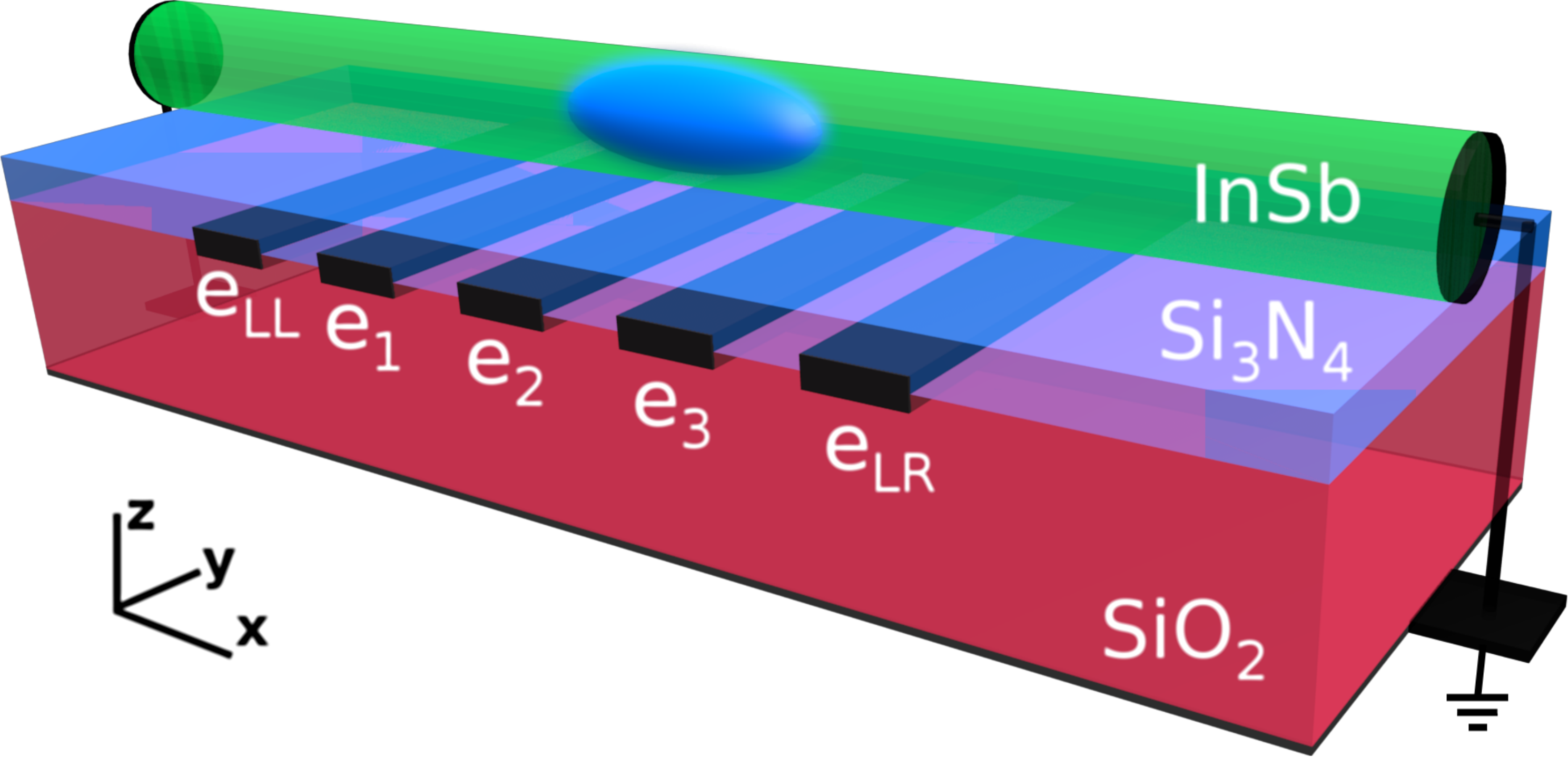}
\caption{\label{fig:1} (color online). The schematic view of the nanodevice. The presented structure of the nanowire on top of five
narrow bottom gates similar to those considered in Refs.~[\onlinecite{1,2,3,4}]. The quantum dot is formed using the middle three gates.
}
\end{figure}
{The two nanodevices discussed below are based on a gated semiconductor nanowire as depicted in Figs.~\ref{fig:1} and~\ref{fig:2}. 
The quantum dot is defined within the nanowire by the local gating in a similar manner as in the experimental set-ups~\cite{1,2,3,4}.}

The entire structure is placed on a strongly doped silicon substrate, covered with a 180~nm layer of $\mathrm{SiO}_2$. 
On the substrates an array of five $40$~nm wide metallic gates are deposited, which are covered with a 20 nm layer of $\mathrm{Si}_3\mathrm{N}_4$. 
The inter-electrode distances is about 40~nm. The InSb nanowire with diameter 50~nm is deposited on the $\mathrm{Si}_3\mathrm{N}_4$ layer. 
The $\mathrm{Si}_3\mathrm{N}_4$ layer isolates the gates from a nanowire\cite{4}.

The gates $\mathrm{e}_1$ and $\mathrm{e}_3$ form tunnel barriers and thus create a quantum dot region in the center of the wire, just above gate $\mathrm{e}_2$. In the proposed set-up the dot is occupied by the single electron. The electrode $\mathrm{e}_2$ controls the electron confinement in a $z$ direction (perpendicular to the structure layers). The positive voltages are applied to the two outer gates, $\mathrm{e}_{LL}$ and $\mathrm{e}_{LR}$, in order to create high density of the electron gas in the dot leads.

The nanowire is contacted with ohmic electrodes on the ends. Zero voltage is applied to the strongly doped substrate and both wire ends, creating a potential reference level.

\begin{figure}[tb!]
\includegraphics[width=9cm]{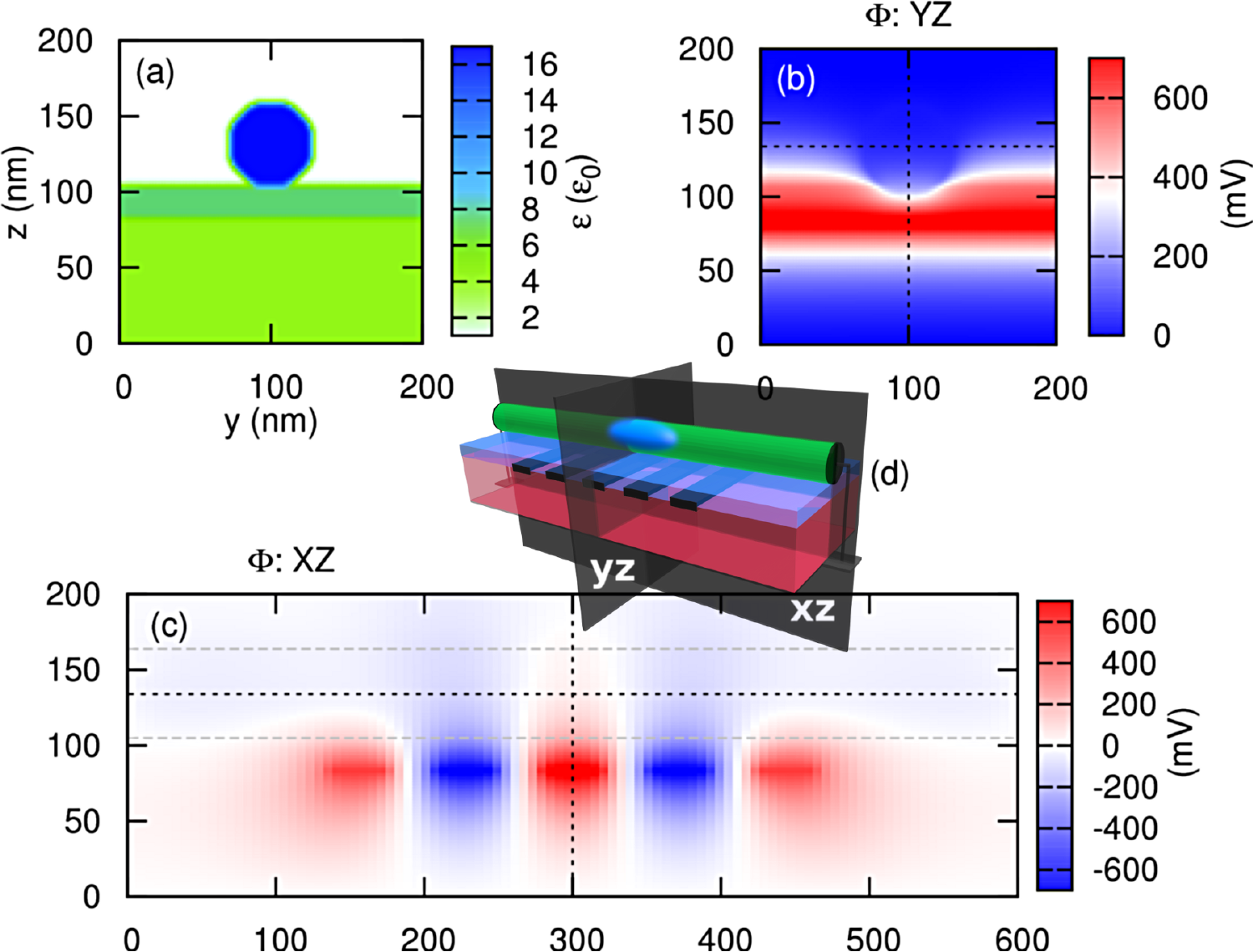}
\caption{\label{fig:3} (color online). The electric permittivity takes different values for each device element. (a) Map of the permittivity values on surface perpendicular to the wire direction.
An example of the $\Phi$ potential distribution in the device on the surface $\mathrm{YZ}$ (b) and $\mathrm{XZ}$ (c). Surfaces marked on the inset (d).}
\end{figure}

The proposed electron spin manipulation scheme is all-electrically controlled. Thus in order to describe the proposed device properly, we need to find the electric potential distribution in the whole system. The studied structure is composed of the materials with different electric permittivities. Furthermore there are various sources of the electric charge: single electron in quantum dot, electron gas in the quantum wire outside the quantum dot region, and charge induced on the material interfaces (permittivity discontinuities) and on the surface of the electrodes. Moreover, one has to include proper boundary conditions and voltage applied to the electrodes. This all affects the electrostatic potential distribution in the nanodevice. In our work it is modeled by a generalized form of the 3D Poisson equation: 
\begin{equation}
\boldsymbol\nabla \cdot \left(\epsilon_0\epsilon(\mathbf{r})\boldsymbol\nabla \Phi(\mathbf{r},t)\right)=-\left(\rho_\mathrm{e}(\mathbf{r},t)+\rho_\mathrm{eg}(\mathbf{r},t)\right),
\label{poiss1}
\end{equation}
where $\rho_\mathrm{e}$ and $\rho_\mathrm{eg}$ is the charge density of a single electron in a quantum dot and electron gas confined within the wire, respectively. 
The form of the equation (\ref{poiss1}) takes into account the fact that the electric permittivity $\epsilon(\mathbf{r})$ is spatially dependent and has different values for each material---see Fig.~\ref{fig:3}(a). 
The Poisson equation (\ref{poiss1}) is numerically solved on a 3D grid that covers the whole nanodevice.
We have applied a central finite difference method for calculating both potential and dielectric function gradient in the left side of Eq. (\ref{poiss1}). 
To increase the  the stability of the method, the function describing the $\epsilon(\mathbf{r})$  is blurred in the border areas between different materials. 
An example of the $\Phi$ potential distribution, which is a solution of the (\ref{poiss1}) equation, can be found in Fig.~\ref{fig:3}. 
We see that the key factors generating the potential distribution are voltages applied to the electrodes and permittivity distribution in the device.
A short analysis of contribution of the various components to the electron confinement potential shape will be discussed in the appendix.

The two outer electrodes, $\mathrm{e}_{LL}$ and $\mathrm{e}_{LR}$, control the density of the electron gas on both sides of the wire, outside the quantum dot region.
We assume that the electron gas density is nonzero wherever the local potential energy $-|e|\Phi(\mathbf{r},t)$ is below the electron gas Fermi energy $E_\mathrm{F}$ within the wire. 
Moreover, due to Coulomb blockade, the electron gas does not enter into the area of the dot confinement potential.
In the remaining area, in the region where the local potential energy is below the Fermi level, we assume that\cite{31}
\begin{equation}
\rho_\mathrm{eg}(\mathbf{r},t)=-|e|\frac{2\sqrt{2}}{3\hbar^3\pi^2}((|e|\Phi(\mathbf{r},t)+E_\mathrm{F})\,m^{\ast})^{3/2} \times w(y,z),
\label{eg}
\end{equation}
which means that charge density in the right side of Eq.~(\ref{poiss1}) depends on total potential itself, which requires the use of a self-consistent numerical method.
The weight function $w(y,z)=(2\pi\sigma)^{-1}\exp[-(y-y_0)^2/2/\sigma^2-(z-z_0)^2/2/\sigma^2]$ 
describes the confinement in the lateral ($y$,$z$) directions.
It is centered at the symmetry axis of the wire (passing through the wire symmetry point $(x_0,y_0,z_0)$ 
and is parallel to the $x$ axis), 
The $\sigma=7$~nm is the dispersion parameter.
As an example, in InSb material for $|e|\Phi+E_\mathrm{F}=100$~meV, we have the electron gas density $\rho_\mathrm{eg}\simeq3\times10^{17}$~cm$^{-3}$. 
The expression from Eq.~(\ref{eg}) can be derived starting from the formula for the free-electron Fermi gas density:
$k_\mathrm{F}^3/3/\pi^2$.\cite{32} 

The Fermi level of the electron gas has been chosen in such a way that the density of the electron gas for electrically isolated (from any charge depleting voltages) wire equals $\rho^0_\mathrm{eg}=5\times10^{16}$~cm$^{-3}$.\cite{51}
We have additionally performed self-consistent calculations of the  concentration of the electron gas within the wire with zero voltages applied to all bottom electrodes, and tuned the value of $E_F=70$~meV 
to achieve average density of the electron gas within the wire equal to $\rho^0_\mathrm{eg}$.

For the Poisson equation, we apply von Neumann boundary conditions:
\begin{equation}
\boldsymbol n \cdot \boldsymbol\nabla \Phi=0,
\label{boundary}
\end{equation}
where $\boldsymbol n$ is a vector normal to the box surface. 
The Neumann conditions are consistent with Gauss' law for charge neutrality of the computational box content\cite{31}.
We use such conditions at the top surface of the computational box, above the wire and at the walls, except in the region of the grounded wire ends that touch the box walls. 
Besides the wire ends, a zero potential is applied to the substrate, which is the bottom boundary. The total potential on the electrodes is constant and is determined by the applied voltages.

The choice of a proper boundary condition is most problematic for the top surface of the computational box, which is the furthest from the substrate (with strict zero potential condition). However, it turns out, that if we change the~(\ref{boundary}) conditions on the ceiling to zero potential---as if we cover the structure by the grounded metal plate, the effective Rashba coupling amplitude will change only by less than 1\%.


In proposed set-up a single electron is trapped in the quantum dot. The time evolution of the spin-$1/2$ electron is described by the time-dependent Schr\"{o}dinger equation 
\begin{equation}
i\hbar\frac{\partial}{\partial t}\Psi(\mathbf{r},t)= H(\mathbf{r},t) \Psi(\mathbf{r},t),
\label{schr}
\end{equation}
where the electron wave function has a two-row form: $\Psi(\mathbf{r},t)=\left(\psi_{\uparrow}(\mathbf{r},t),\psi_{\downarrow}(\mathbf{r},t)\right)^T$, $\mathbf{r}=(x,y,z)$
and the arrow indicates the spin projection onto the quantization axis ($z$).
The Hamiltonian of the considered system is
\begin{equation}
H(\mathbf{r},t)=\left(-\frac{\hbar^2}{2m^{\ast}}\boldsymbol\nabla^2+\varphi(\mathbf{r},t)\right)\!1_2+H_R(\mathbf{r},t),
\label{ham}
\end{equation}
where $1_2$ is a $2\times2$ identity matrix, $m^{\ast} = 0.014m_0$ is the effective mass of the electron in the InSb material ($m_0$ is the
free electron mass). The Hamiltonian (\ref{ham}) is time-dependent due to the second and third component, which are calculated based on the $\Phi(\mathbf{r},t)$ total electrostatic potential distribution.

The $\Phi(\mathbf{r},t)$ varies in time within the device due to the voltage changes applied to the electrodes and a non stationary charge distribution of the single electron in a quantum dot $\rho_\mathrm{e}(\mathbf{r},t)$ and the electron gas  $\rho_\mathrm{eg}(\mathbf{r},t)$.
On the other hand, the charge density of a single electron $\rho_\mathrm{e}(\mathbf{r},t)$ depends on the variable potential $\varphi(\mathbf{r},t)$, and is calculated from the electron wave function: $\rho_\mathrm{e}(\mathbf{r},t)=-|e|\Psi^\dag(\mathbf{r},t)\Psi(\mathbf{r},t)$.
To take into account these dependencies, the Schr\"{o}dinger equation is solved numerically in an iterative manner with the Poisson equation
solved at every time step of the iteration procedure. The Poisson method itself is self-consistent due to the fact that the electron gas distribution $\rho_\mathrm{eg}(\mathbf{r},t)$ depends on the potential $\Phi(\mathbf{r},t)$.

The electron confinement potential $\varphi(\mathbf{r},t)$ should not contain electron self-interaction potential $\varphi_e(\mathbf{r},t)$, 
thus we subtract the influence of the electron: $\varphi'(\mathbf{r},t)=\Phi(\mathbf{r},t)-\varphi_e(\mathbf{r},t)$.\cite{15}
Moreover, to take into account the conduction band offset at the wire/dielectric interface, 
we subtract a constant level everywhere outside the wire: $\varphi(\mathbf{r},t)=\varphi'(\mathbf{r},t)-V_\mathrm{offset}$, 
and $\varphi(\mathbf{r},t)=\varphi'(\mathbf{r},t)$ within the wire. The value of InSb/$\mathrm{Si}_3\mathrm{N}_4$ conduction band offset is $2.4$~eV,\cite{34} while for the InSb/$vacuum$ interface it is higher, and equals $4.6$~eV.\cite{34} 

To simplify calculations and keep confinement plots transparent, we assume single offset value as $V_\mathrm{offset}=1.0$~V.
Taking the higher offset value does not significantly change the shape of the electron charge distribution $\rho_\mathrm{e}$.

The single-electron self-interaction potential is calculated from the equation 
\begin{equation}
\boldsymbol\nabla^2 \varphi_e(\mathbf{r},t)= -\frac{\rho_\mathrm{e}(\mathbf{r},t)}{\epsilon_0\epsilon(\mathbf{r})},
\label{poiss2}
\end{equation}
with the exact boundary conditions calculated directly from the Coulomb law, by integrating over the electron charge distribution in the system.
This method allows us to calculate $\varphi'$ (also $\varphi$) where we take into account voltages applied to the electrodes, charge distribution of the electron gas and also the charge induced on the interfaces between different materials and electrodes.


The non uniform electric field $\mathbf{E}(\mathbf{r},t)$ within the wire is the source of the spatially inhomogeneous Rashba SOI coupling\cite{hanson,winkler}: 
\begin{equation}
H_R(\mathbf{r},t)=\alpha_\mathrm{3D} e \left(\mathbf{E}(\mathbf{r},t)\times\mathbf{k}\right)\cdot\boldsymbol\sigma.
\label{rashba}
\end{equation}
The electric field $\mathbf{E}(\mathbf{r},t)=-\boldsymbol\nabla\varphi'(\mathbf{r},t)$ is calculated for the potential $\varphi'$ without inclusion of the conduction band offset at the wire/dielectric interface.\cite{9} The wave vector is $\mathbf{k}=-i\boldsymbol\nabla$,
$\boldsymbol\nabla\equiv\left[\partial_x,\partial_y,\partial_z\right]$. 
The vector $\boldsymbol\sigma$ is defined by Pauli matrices: $\boldsymbol\sigma\equiv\left[\sigma_x,\sigma_y,\sigma_z\right]$. The InSb Rashba SOI coefficient is 
$\alpha_\mathrm{3D}=5.23$~nm$^2$.\cite{winkler}
{We assume that nanowire is grown along [111] crystallographic direction which allows us to neglect the Dresselhaus SOI.}\cite{winkler}

\section{RESULTS AND DISCUSSION}

In the next section we will discuss the $\varphi(\mathbf{r},t)$ potential distribution in the nanodevice and propose a scheme where the potential $\varphi(\mathbf{r},t)$ is changed by modulation of the voltages applied to the electrodes. This changes in turns lead to an oscillatory behavior of  confinement potential of the electron and consequently to a modulation of the Rashba SOI. We show that the suitable changes of the RSOI with simultaneous oscillating motion of the confined electron leads to an effective spin rotations.

\subsection{Nanodevice for the $y$-axis spin rotations}
\begin{figure}[bt]
\includegraphics[angle=-90,width=8.7cm]{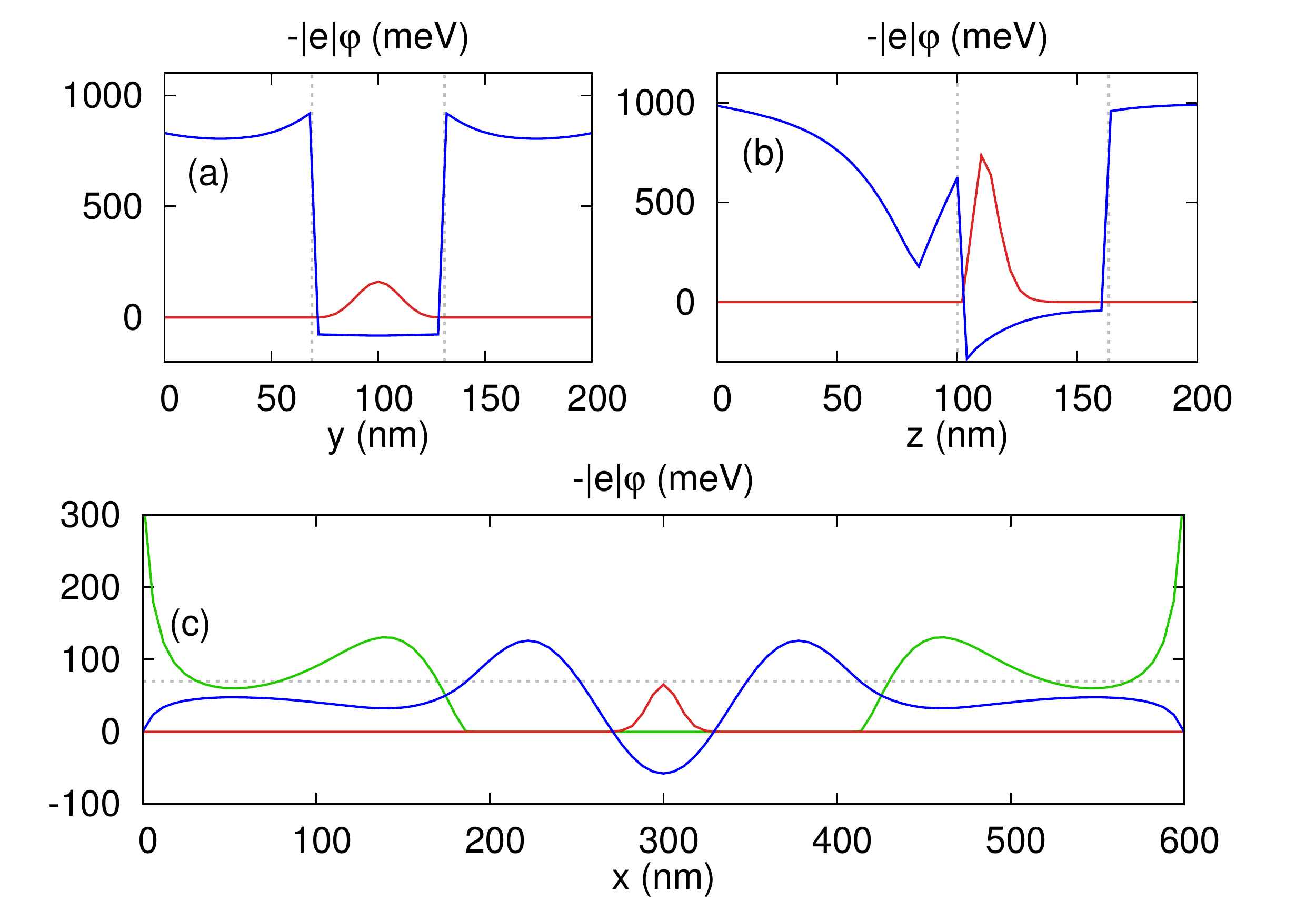}
\caption{\label{fig:4} (color online). The electron confinement potential $\varphi(\mathbf{r})$ marked by blue curves plotted along the axes denoted in Fig.~\ref{fig:3} by dashed black lines 
(all passing through the wire symmetry point $(x_0,y_0,z_0)$ and parallel to the system axes).
In this way we can illustrate the confinement in all three directions $y$, $z$, $x$---(a), (b), (c) subfigures respectively. 
Red curves illustrate the electron density $\rho_\mathrm{e}(\mathbf{r})$ (arb. unit), green one, the electron gas density $\rho_\mathrm{eg}(\mathbf{r})$ (arb. unit) within the wire. 
Both are plotted along the same axes as the confinement potential.}
\end{figure}

First let us consider the operation of the device from Fig.~\ref{fig:1}.
At initial time, the voltages applied to the electrodes take the following values: $V_1(t=0)=-750$~mV, $V_2(t=0)=800$~mV, $V_3(t=0)=-750$~mV, $V_{LL}=V_{LR}=550$~mV. 
This creates the initial electron confinement within the device depicted in Fig.~\ref{fig:4}.
The central gates $\mathrm{e}_1$, $\mathrm{e}_2$, and $\mathrm{e}_3$ that are underneath the InSb wire create confinement along the wire, in $x$ direction---see Fig.~\ref{fig:4}(c). 
Moreover, gate $\mathrm{e}_2$ is a source of potential asymmetry in the $z$ direction---see Fig.~\ref{fig:4}(b). 
The conduction band offset at the wire interface is visible in the lateral ($y$,$z$) confinements---see Fig.~\ref{fig:4}(a,b).

\begin{figure}[htb!]
\includegraphics[angle=-90,width=8.7cm]{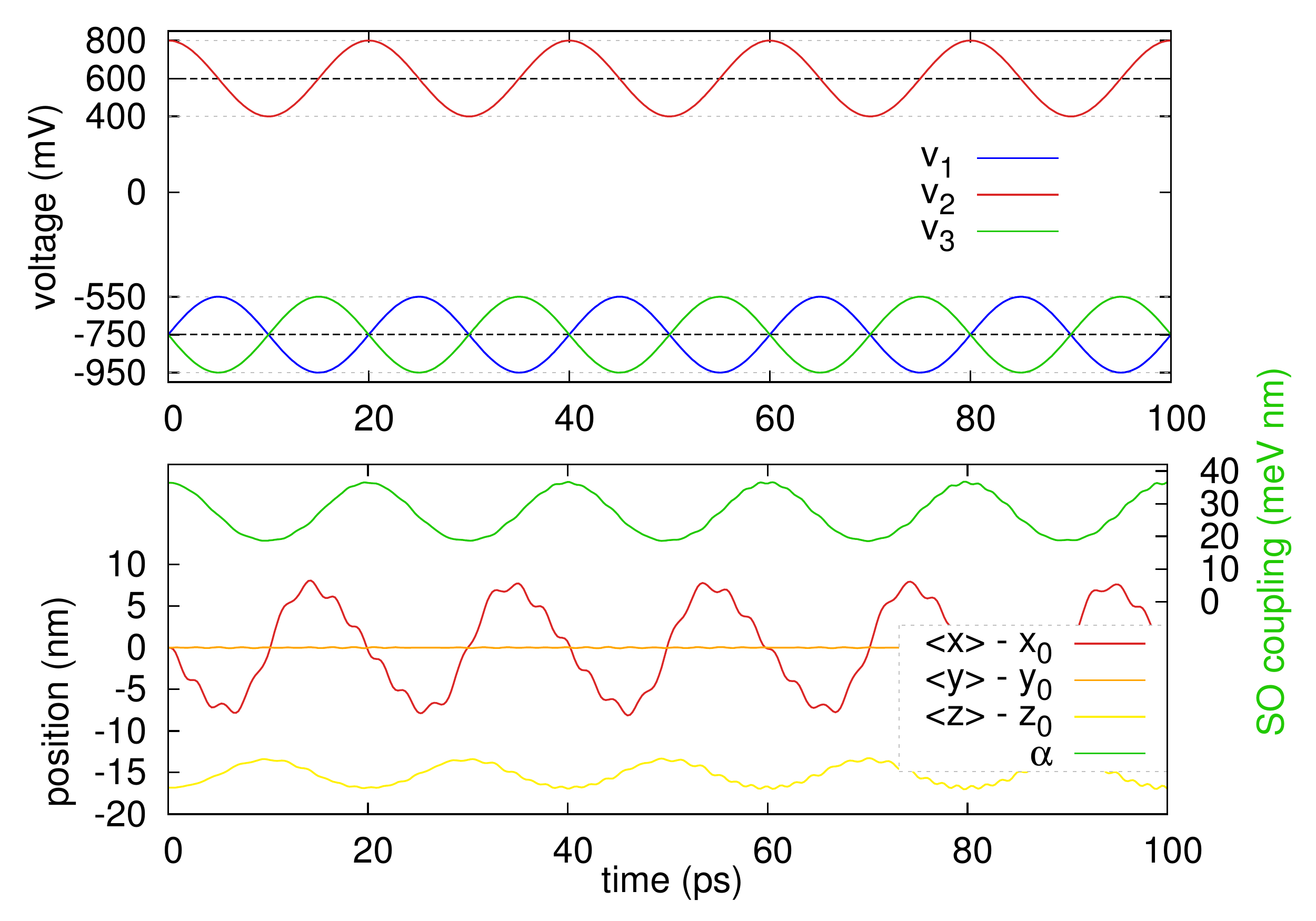}
\caption{\label{fig:7} (color online). (top) The voltages applied to the central gates $\mathrm{e}_1$, $\mathrm{e}_2$, $\mathrm{e}_3$ 
during the first 100~ps are marked by red, green and blue color for $V_{1}$, $V_{2}$, $V_{3}$ respectively. 
(bottom) The expectation values of the electron position $\langle x \rangle$, $\langle y \rangle$, $\langle z \rangle$ relatively to the wire symmetry point $(x_0,y_0,z_0)$ 
marked by red, orange and yellow curve respectively. Estimated value of the Rashba spin-orbit coupling marked by green curve.}
\end{figure}

Additionally, by modulating the voltages applied to central gates $\mathrm{e}_1$, $\mathrm{e}_2$, $\mathrm{e}_3$ 
we can induce motion of an electron ($\mathrm{e}_1$ and $\mathrm{e}_3$) and oscillating SOI coupling ($\mathrm{e}_2$).
Let us check what happens if the voltages are being changed in oscillatory way over time in a following way:
$V_{1}(t)=V_{10}+V_{11}\sin(\omega t)$, $V_{3}(t)=V_{30}-V_{31}\sin(\omega t)$, $V_{2}(t)=V_{20}+V_{21}\cos(\omega t)$, with the offsets $V_{10}=V_{30}=-750$~mV, $V_{20}=600$~mV, and the amplitudes $V_{11}=V_{21}=V_{31}=200$~mV.
The evolution of the voltage levels during the first 100~ps are depicted in Fig.~\ref{fig:7}(top).
{We set the voltages oscillation frequency to $\omega/2\pi = 50$~GHz.}
\begin{figure}[h]
\includegraphics[angle=-90,width=8.7cm]{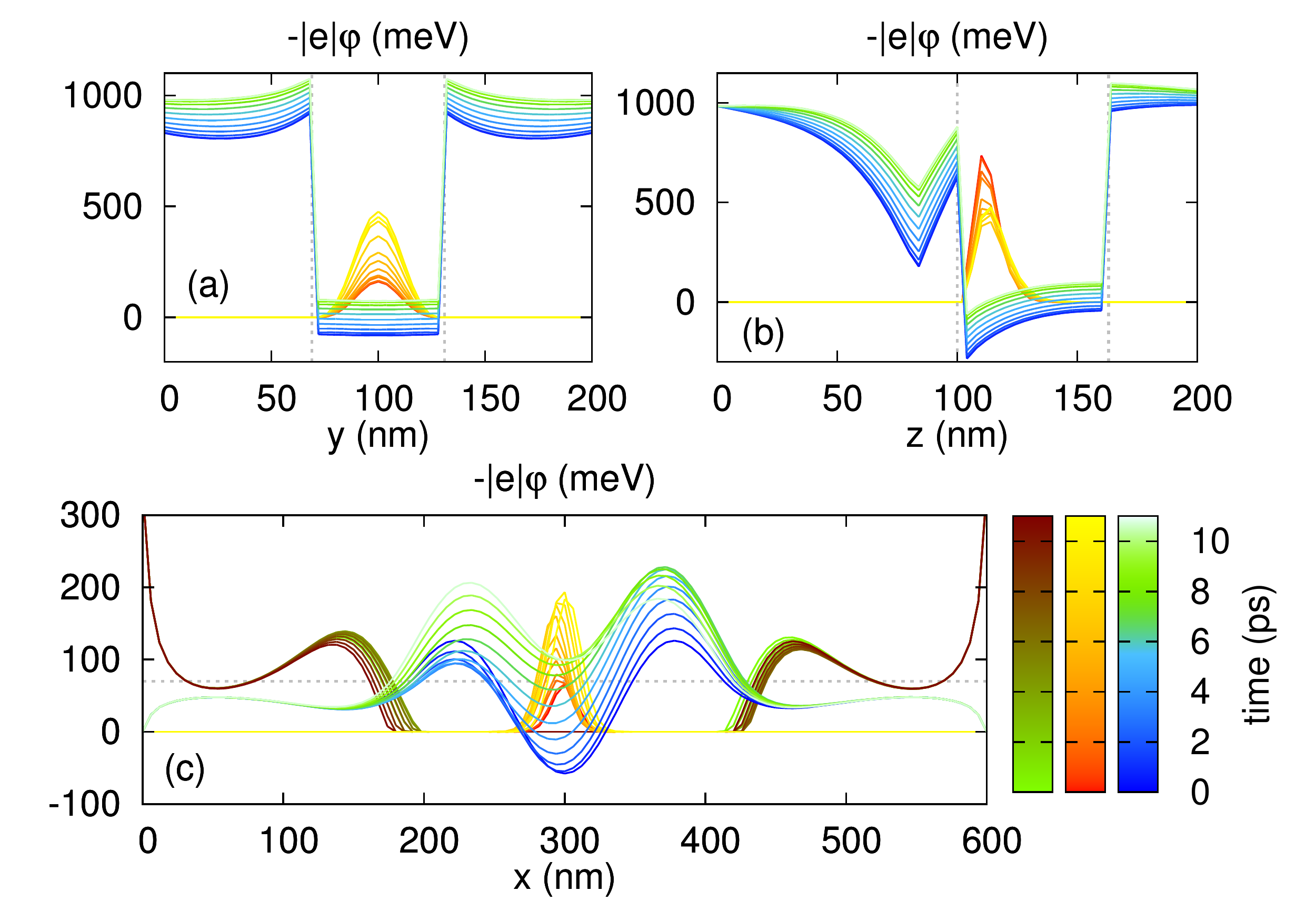}
\caption{\label{fig:5} (color online). The electron confinement potential (blue to green palette), the electron density (red to yellow) and the electron gas density (green to brown) shapes 
during the first half of the voltage change period, i.e., $t=0$ - $10$~ps (presented same as in Fig.~\ref{fig:4}).}
\end{figure}
\begin{figure}[b]
\includegraphics[angle=-90,width=8.7cm]{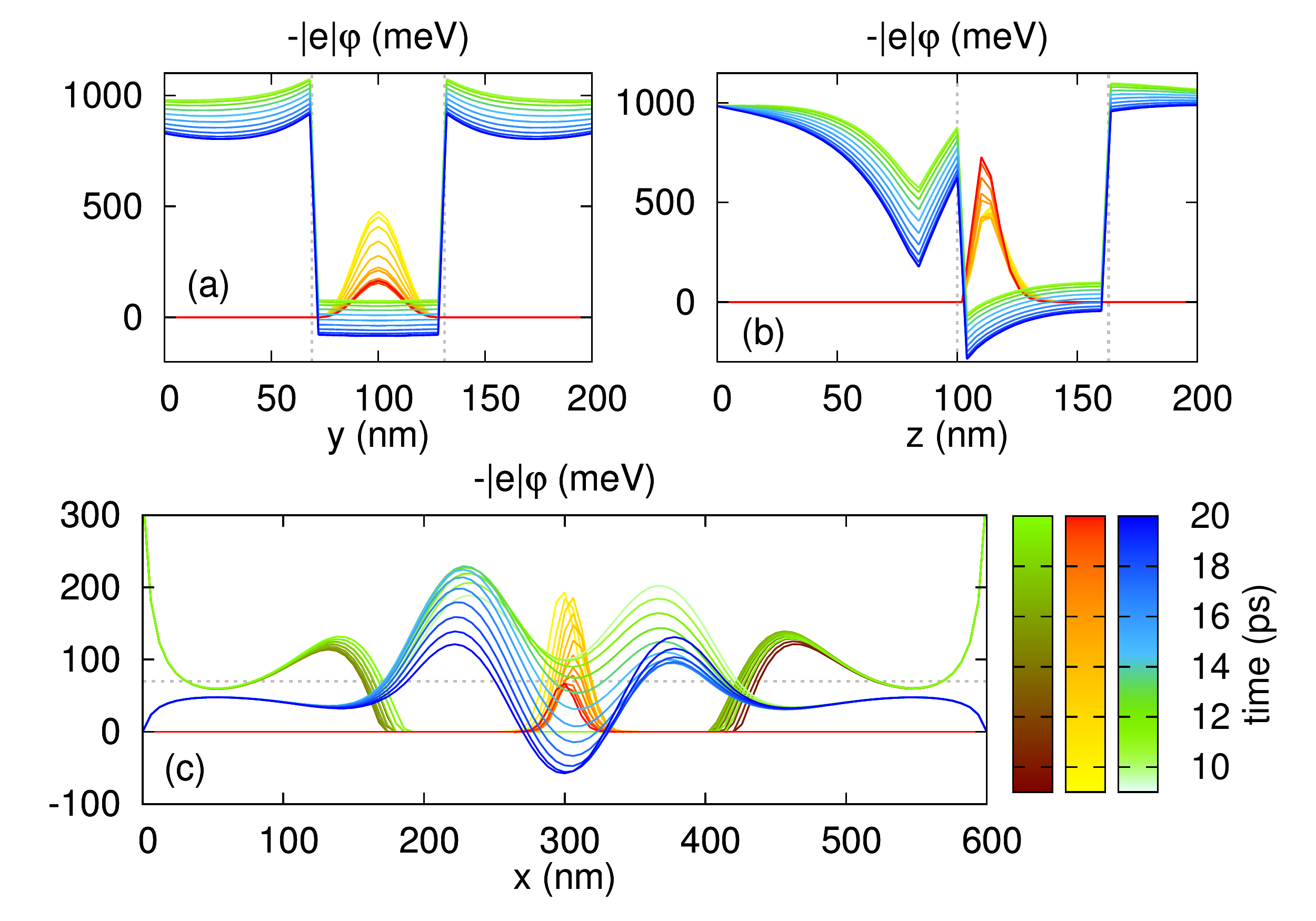}
\caption{\label{fig:6} (color online). The electron confinement potential (green to blue), the electron density (yellow to red) and the electron gas density (brown to green) shapes
during the second half of the voltage changes period, i.e., $t=10$ - $20$~ps.}
\end{figure}

By changing the voltage one can modify confinement potential for the electron. Evolution of the confinement potential shapes during the first 20~ps---period of voltage changes are depicted in Fig.~\ref{fig:5} (for $t=0$ - $10$~ps) and Fig.~\ref{fig:6} (for $t=10$ - $20$~ps).
Changes in $V_{1}(t)$ and $V_{3}(t)$ voltages generate motion of the potential well in which the electron is trapped, which in turns cause an oscillatory motion of the electron itself.
In the Fig.~\ref{fig:7}(bottom) we can see expectation values of the electron position, and observe oscillatory movement of the electron along the $x$ direction. 
Moreover, changes of the $V_{2}(t)$ voltage generate a modulation of the slope of the asymmetric potential in the $z$ direction,
visible in Figs.~\ref{fig:5}(b) and \ref{fig:6}(b).
This leads to an oscillatory behavior of the Rashba SOI. 
In the Hamiltonian $H_R$---Eq.~(\ref{rashba}) the 3D RSOI coupling is calculated exactly from the electric field. However, in order to estimate the effective value of the SOI coupling and illustrate its oscillatory behavior we use a following formula
$\alpha=\alpha_\mathrm{3D} e \sqrt{\langle E_x \rangle^2 + \langle E_y \rangle^2 + \langle E_z \rangle^2}$, where the mean values of the electric field components are averaged over the actual electron probability distribution. Values of the $\alpha(t)$ for the first 100~ps are depicted at the bottom part of the Fig.~\ref{fig:7}(green curve).

\begin{figure}[t]
\includegraphics[width=8.7cm]{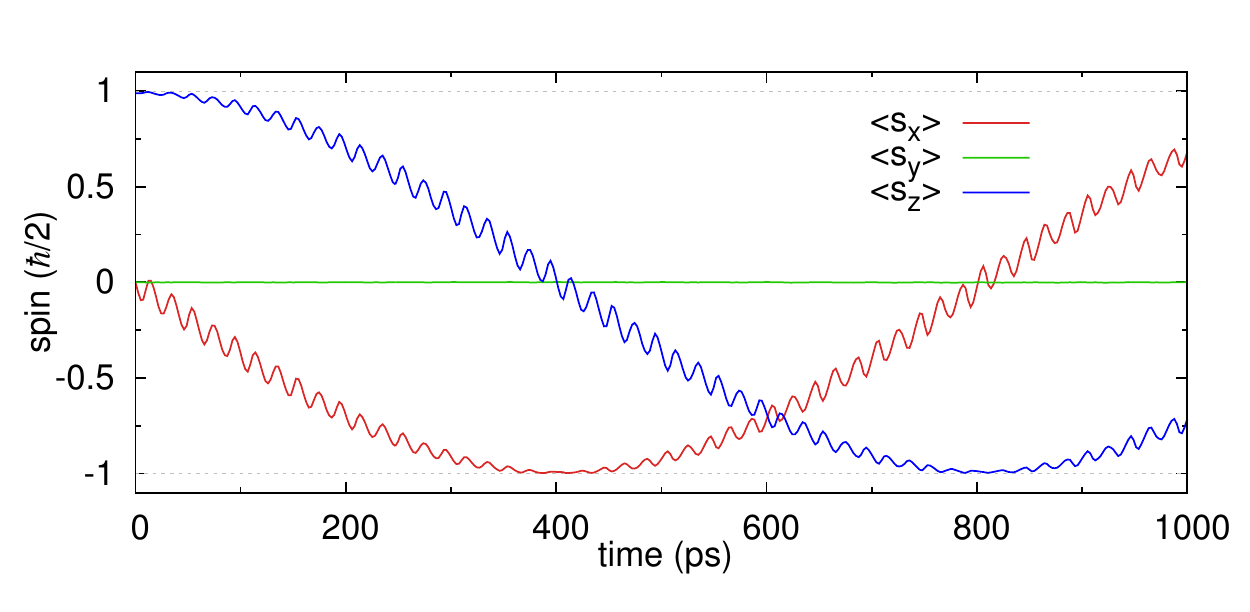}\\
\vspace{-4mm}
\includegraphics[width=4.5cm]{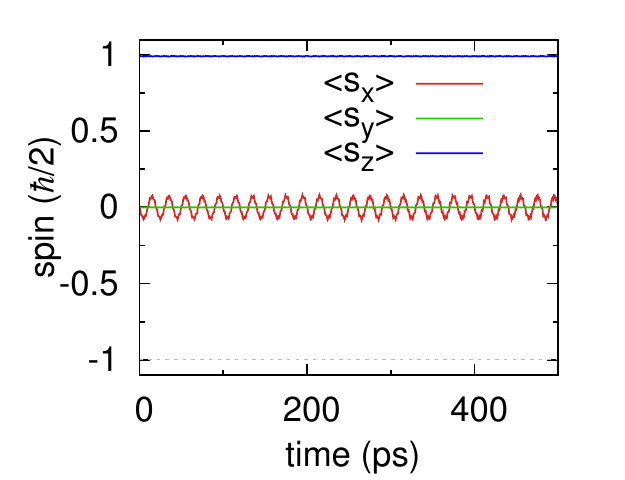}
\hspace{-6mm}
\includegraphics[width=4.5cm]{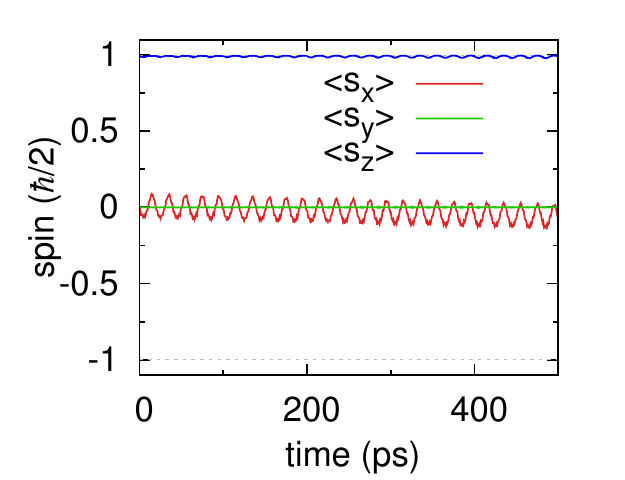}
\caption{\label{fig:8} (color online). (top) The expectation values of the electron spin components during the device operation for the voltage modulation presented in Fig.~\ref{fig:7}(top). 
(bottom) Similar, but now the $V_2$ is constant over time (left), or $V_2$ is in phase with $V_1$ and $V_3$ (right).}
\end{figure}
The electron moves back and forth along the $x$-axis with simultaneous changes of the Rashba coupling $\alpha(t)$ which has a different value for the electron moving in "$+x$" and "$-x$" directions. This leads to the effective spin rotation as shown in Fig.~\ref{fig:8}(top). 
Initially, the electron spin is aligned along the $z$ spin axis. Than due to the oscillatory motion of the electron in the non-homogeneous and non stationary Rashba SOI field, the spin rotates around $y$-axis. After approximately 400~ps the $\pi/2$ rotation is accomplished and the spin is now oriented along the $x$ spin axis. To obtain such effect, it is necessary to shift the phase by $\pi/2$ between the oscillations of voltages $V_2(t)$ and $V_1(t)$ or $V_3(t)$. 
For the constant (in time) value of the Rashba SOI coupling (no modulations of the voltage on $\mathrm{e}_2$, i.e. $V_2=V_\mathrm{20}$) the spin components are only slightly affected by oscillatory motion of the electron, but without effective spin rotation. This is illustrated on the Fig.~\ref{fig:8}(bottom, left). Similarly, we do not get effective spin rotation if the voltage $V_2(t)$ is in phase with the voltages $V_1(t)$ or $V_3(t)$, i.e., $V_{2}(t)=V_{20}+V_{21}\sin(\omega t)$. Results of our simulations for such a case can be found on  the Fig.~\ref{fig:8}(bottom, right).

The voltages are chosen in such a way, that the heights of the tunnel barriers ($V_1$, $V_3$) and the confinement depth ($V_2$) inside the dot are such, that no electron tunnel in or out the dot region, during the simulation. Whereas at the device initialization stage one should lower the barriers $V_1$,$V_3$ and select $V_2$ to be slightly below the threshold above which the last electron escapes the dot. This means that we are right behind the entrance to the first Coulomb diamond on the charge stability diagram. This procedure initialize the dot with exactly a single active electron. Then we raise back voltages on the electrodes $\mathrm{e}_1$ and $\mathrm{e}_3$, and then lower voltage on the electrode $\mathrm{e}_2$ to the offset levels: $V_1=V_{10}$, $V_3=V_{30}$ and $V_2=V_{20}\!+\!V_{21}$.

In all the presented simulations, the frequency $\omega/2\pi$ of the voltages oscillations is set to $\omega/2\pi=50$~GHz.
But further tests shows that this frequency can be increased up to the $\omega/2\pi\approx 1$~THz value, which is the adiabaticity limit in considered system. Above this threshold, the electron starts to accumulate energy in every cycle of the confinement potential changes.
The presented operation times of the electron spin are about $800$~ps for the spin NOT gate---see Fig.~\ref{fig:8}(top), and are much shorter than the electron spin coherence times in InSb material.\cite{13}
These operation times might be further improved (reduced) by increasing the frequency $\omega/2\pi$.

The proposed electron spin rotation process can be understood as follows.
The electron position oscillations within the wire (mainly along the $x$-axis) shown in Fig.~\ref{fig:7}(bottom panel, red curve)  
with simultaneous varying the SOI (green curve), however not in the phase with the electron oscillation, result in the following spin rotations: During movement in the "$+x$" direction (for about $\delta x\!\sim\!15$~nm that lasts half cycle of the voltage oscillations $\pi/\omega = \delta t=10$~ns) the electron feels some effective spin-orbit field oriented nearly along the $y$-axis which leads to spin rotation. During the return movement "$-x$" it feels effective field with higher amplitude and opposite direction, which consequently leads to opposite spin rotation. Second rotation is bigger due to stronger field at this time, so this both rotations do not cancel and effectively, the electron returns to initial position with rotated spin. Effective field connected with this composite rotation have amplitude which leads to energy splitting, derived from the Eq.~(\ref{rashba}), of $\Delta E_{SO} = \Delta_{SO}\,k/\sqrt{2}$, proportional to spin-orbit coupling difference $\Delta_{SO}$, where we estimate electron momentum $\hbar k$ as $m^\ast \delta x/\delta t$. The factor $1/\sqrt{2}$ results from the fact that we take 
a time-averaged spin-orbit difference, here (see Fig.~\ref{fig:7}(bottom), green curve) we have $\Delta_{SO}\sim20 $~meV~nm. The spin rotations frequency connected to this splitting $\Delta E_{SO}/h$ gives, after inserting the numerical values, $\sim625$~MHz, which sets the timescale $\sim1.6$~ns for the spin rotations seen in Fig.~\ref{fig:8}(top).

The spin rotation performance for given $\omega/2\pi$ is proportional to the electron position oscillation amplitude $\delta x\!/2$ and changes of the spin-orbit coupling $\Delta_{SO}$, which in turn are dependent on the voltage oscillations amplitudes. However, this amplitudes should not be too large, that during the oscillating movement ($V_{11}$ and $V_{31}$) and confinement potential asymmetry modulation ($V_{21}$, or $V_L$,$V_R$ in the second device version) the electron does not tunnel outside the dot.

\subsection{Nanodevice for the $y$ and $z$-axis spin rotations}
\begin{figure}[bth]
\includegraphics[width=6cm]{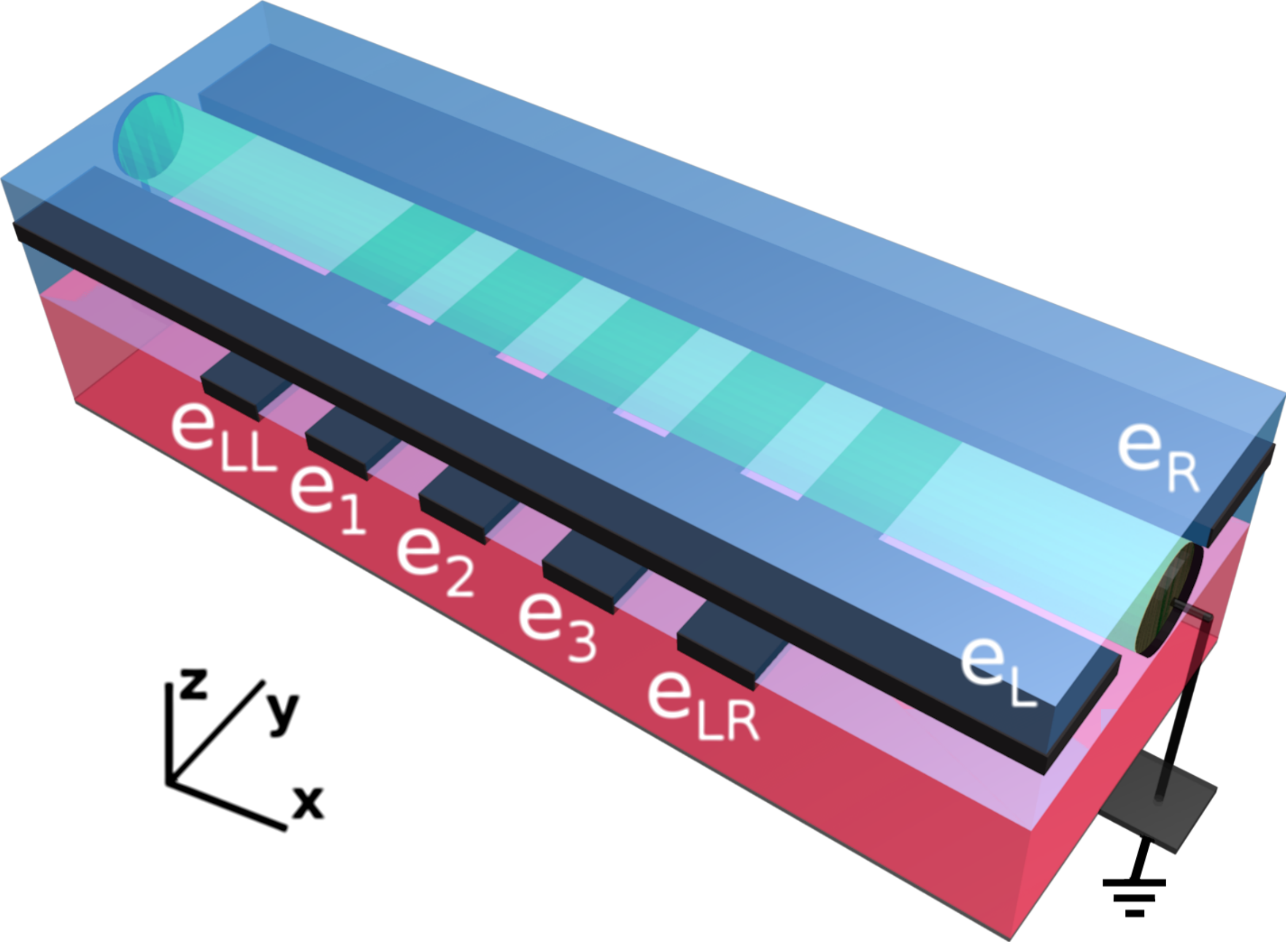}
\caption{\label{fig:2} (color online). The schematic view of the modified version of the nanodevice from previous section A. which is now capable to realize electron spin rotations around $y$ and $z$ axis.
Here we added two electrodes $\mathrm{e}_L$ and $\mathrm{e}_R$ located on both sides of the wire, which allow modification of the electron confinement in the $y$ direction.}
\end{figure}
\begin{figure}[h]
\includegraphics[angle=-90,width=8.7cm]{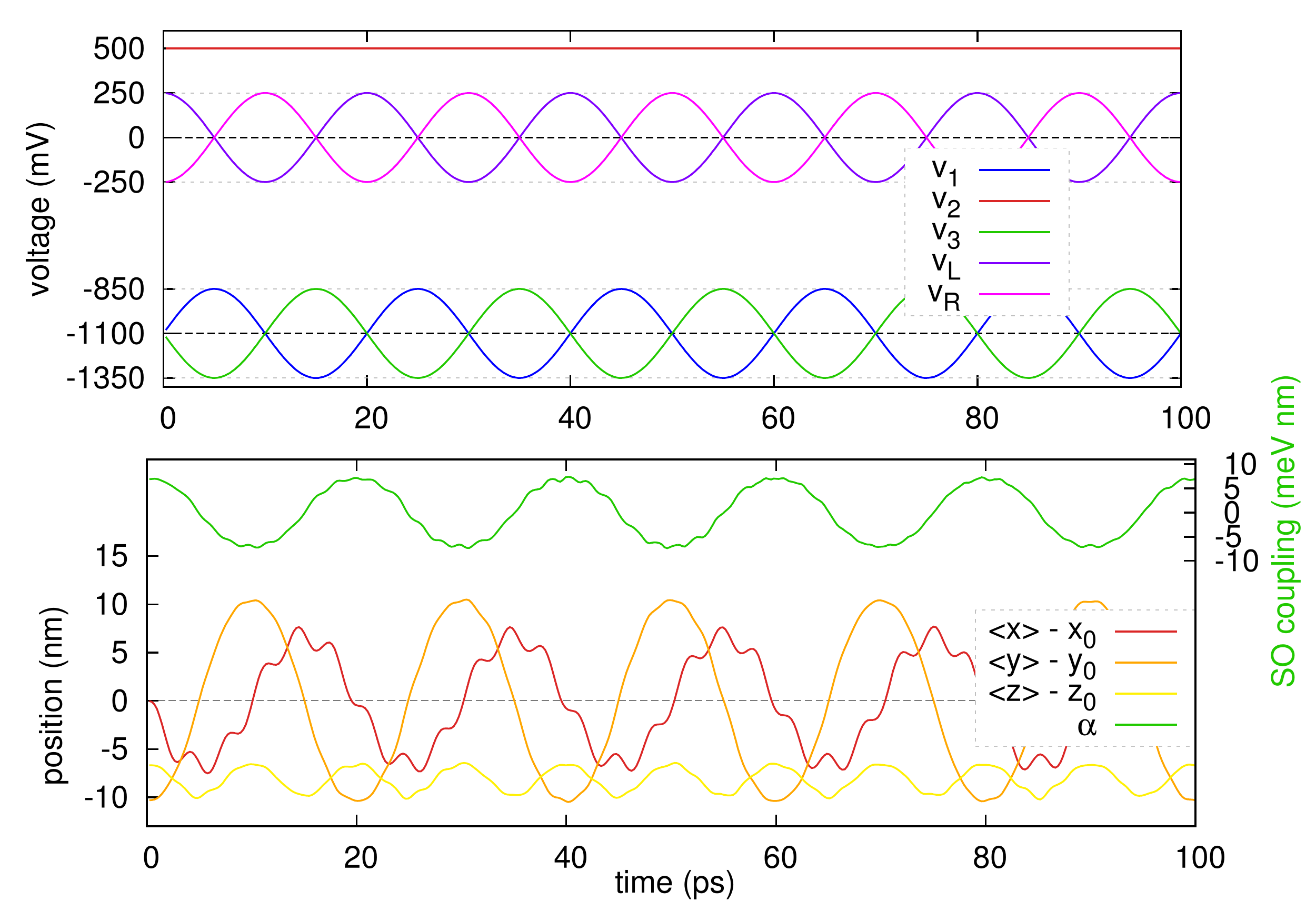}
\caption{\label{fig:12} (color online). Same as Fig.~\ref{fig:7}, but for the device presented in Fig.~\ref{fig:2}.}
\end{figure}

 
Nanodevice proposed in previous section A (depicted in Fig.~\ref{fig:1}) is capable to realize spin rotations around $y$ axis, as presented in Fig.~\ref{fig:8}(top).
Here, we propose a modified version of this device which allows to perform spin rotations on the Bloch sphere around two orthogonal axes separately, 
and thus achieve any single qubit operation.
For this purpose we add two side electrodes $\mathrm{e}_L$ and $\mathrm{e}_R$, as shown in Fig.~\ref{fig:2} to the previous set up.
The oscillating voltages are applied to the $\mathrm{e}_L$ and $\mathrm{e}_R$ electrodes: $V_L(t)=+V_\mathrm{lr}\cos(\omega t)$ and $V_R(t)=-V_\mathrm{lr}\cos(\omega t)$, with $V_\mathrm{lr}=250$~mV.
The voltage applied to the remaining electrodes is similar as in the set up from section A (in this case $V_{10}=V_{30}=-1100$~mV, $V_{11}=V_{31}=250$~mV), but in the current case the voltage applied to the $\mathrm{e}_2$ electrode is constant $V_2=500$~mV. 
The time evolution of the voltages applied to the electrodes are presented in Fig.~\ref{fig:12}(top).
The frequency of voltage oscillations is set to $\omega/2\pi = 50$~GHz.

\begin{figure}[h]
\includegraphics[angle=-90,width=8.7cm]{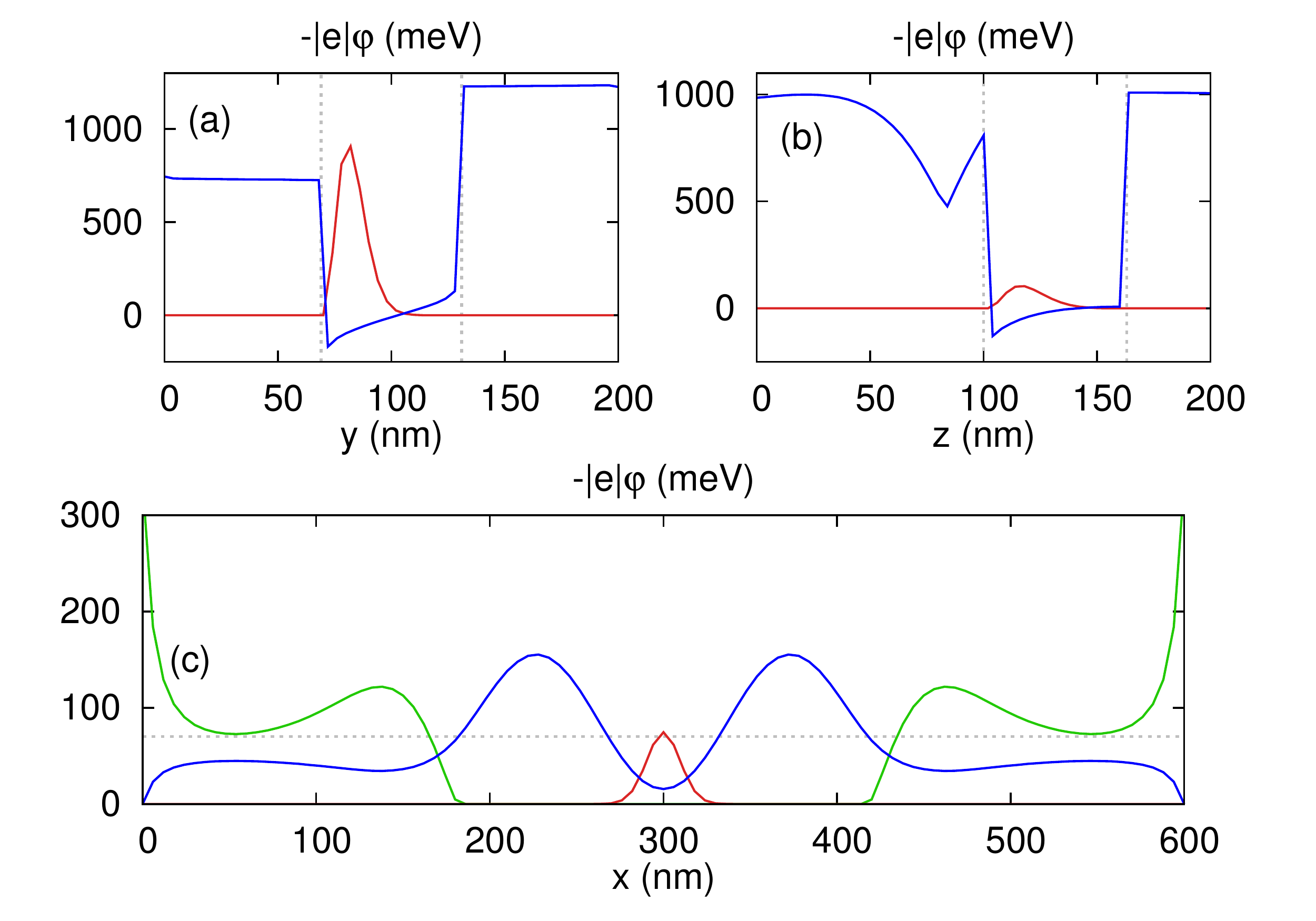}
\caption{\label{fig:9} (color online). Same as Fig.~\ref{fig:4}, but for the device presented in Fig.~\ref{fig:2}.}
\end{figure}
\begin{figure}[h]
\includegraphics[angle=-90,width=8.7cm]{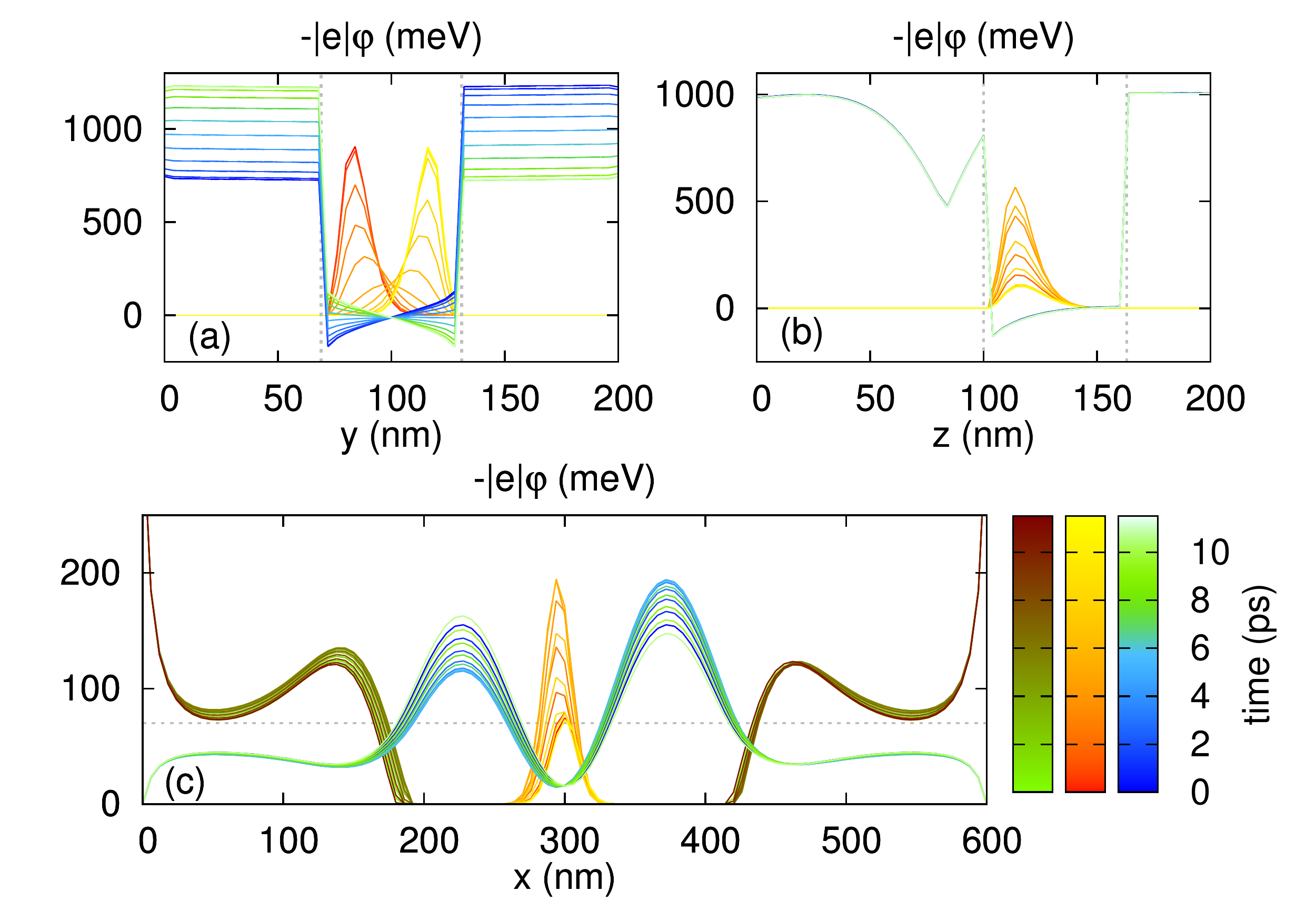}
\caption{\label{fig:10} (color online). Same as Fig.~\ref{fig:5}, but for the device presented in Fig.~\ref{fig:2}.}
\end{figure}
\begin{figure}[h]
\includegraphics[angle=-90,width=8.7cm]{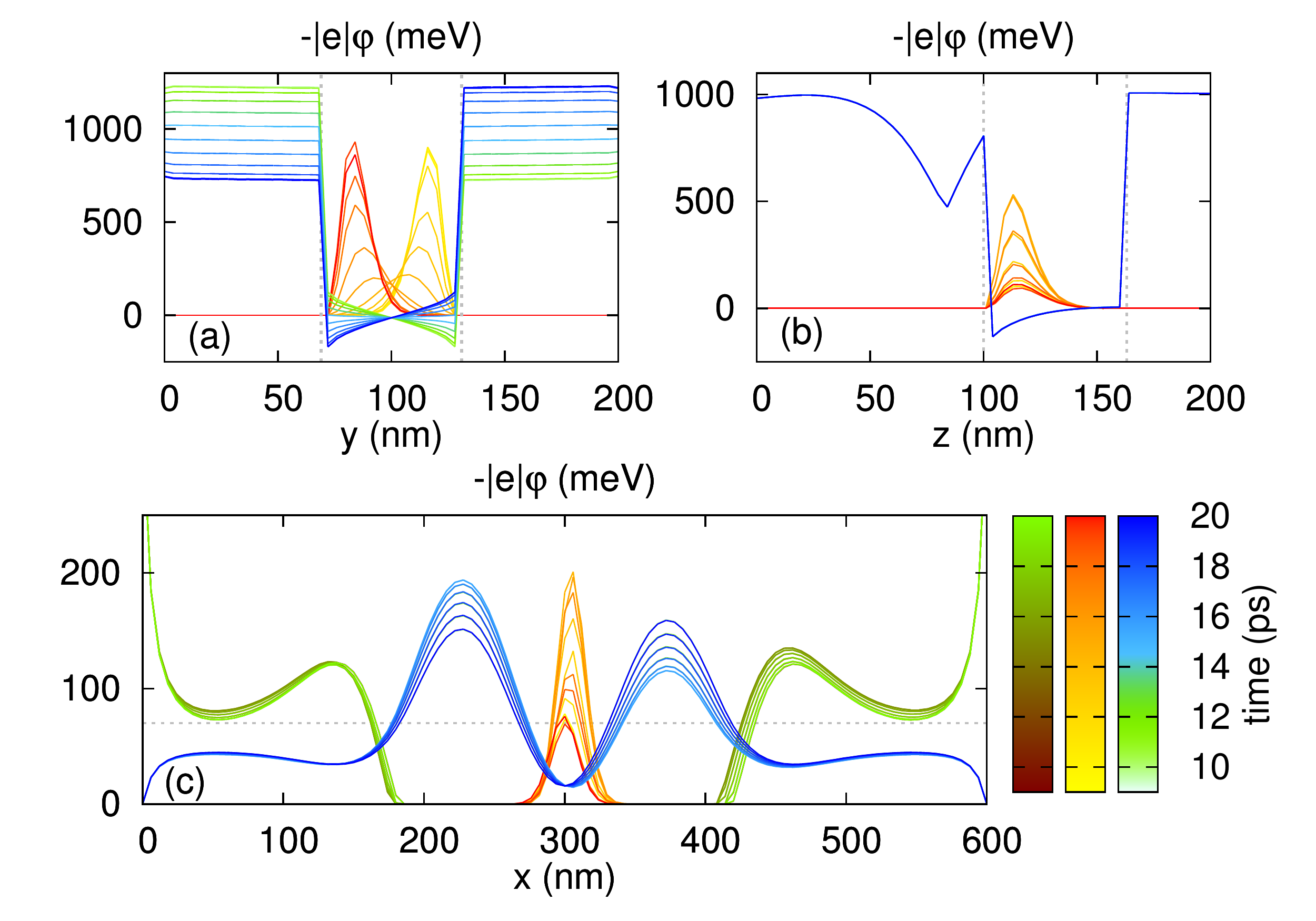}
\caption{\label{fig:11} (color online). Same as Fig.~\ref{fig:6}, but for the device presented in Fig.~\ref{fig:2}.}
\end{figure}

In this case, the voltages $V_L(t)$ and $V_R(t)$ are shifted in phase by $\pi/2$ with respect to voltages $V_1(t)$ or $V_3(t)$ which enforce the oscillating motion of the electron within the wire. The time evolution of the
expectation values of the electron position can be found in Fig.~\ref{fig:12}(bottom). The oscillating voltages create varying confinement potential in the $y$ direction, 
depicted at initial time in Fig.~\ref{fig:9}(a). Its evolution during the first oscillation period of the voltage changes is depicted in Figs.~\ref{fig:10} and \ref{fig:11}.
In a similar way is in previous case (section A.), also here, changes of the slope of the confinement potential, now in the $y$ direction, leads to the modulation of the Rashba SOI.
In order to estimate the varying part of the spin-orbit coupling we use following formula $\alpha=\alpha_\mathrm{3D} e \langle E_y \rangle$.
The time evolution of $\alpha(t)$ which is driven by the oscillating  voltages $V_L(t)$ and $V_R(t)$ is presented in Fig.~\ref{fig:12}(bottom).

\begin{figure}[h]
\includegraphics[width=8.7cm]{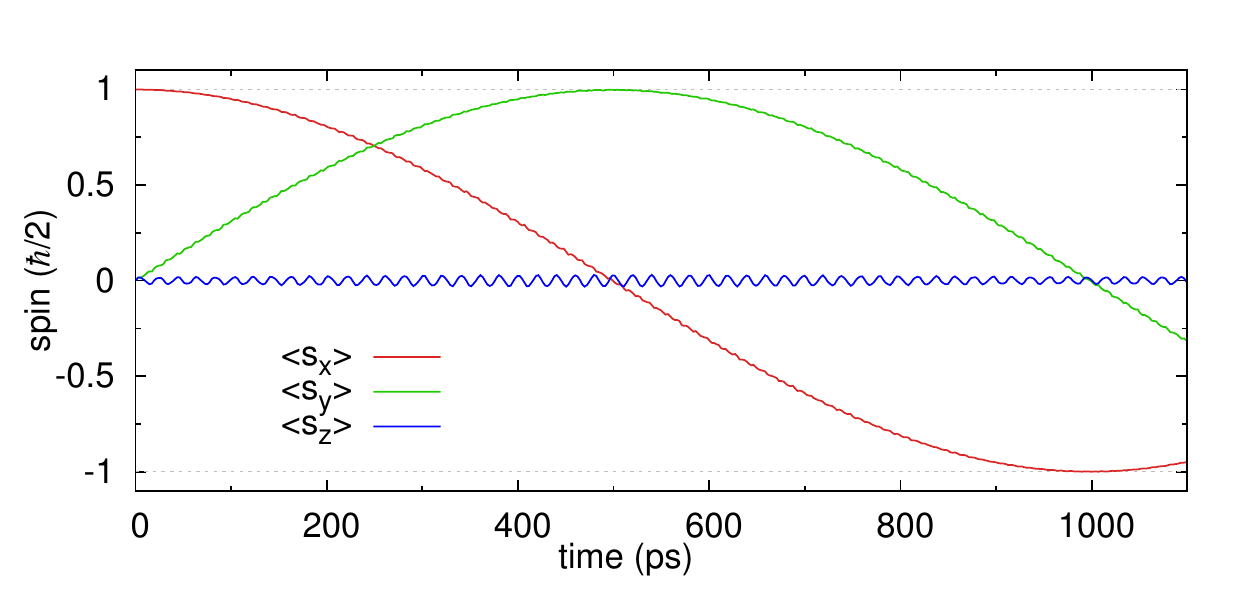}
\caption{\label{fig:13} (color online). Same as Fig.~\ref{fig:8}, but for the device presented in Fig.~\ref{fig:2}.}
\end{figure}

Changes of the Rashba SOI coupling $\alpha(t)$ (generated by the modulation of the confinement potential profile in the $y$ direction), induce the electron spin rotations around the $z$ axis, which are presented in Fig.~\ref{fig:13}. If, in addition, the voltage $V_2(t)$ on the electrode $\mathrm{e}_2$ is modulated as in the first version of the device, one can obtain rotation of the spin independently about two orthogonal axes, here over $y$ and $z$ spin axes.

{The spin-orbit coupling difference on Fig.~\ref{fig:12} (green curve) is approximately $\Delta_{SO}\sim15$~meV~nm. This gives the spin rotations period derived from the effective spin-orbit field $\sqrt{2}h/(\Delta_{SO}\,k)\sim2.1$~ns, which is in agreement with the rotation period shown in Fig.~\ref{fig:13}.}

\section{SUMMARY AND CONCLUSIONS}
In conclusion, we have proposed and simulated operation of nanodevice that can perform single-spin operations in gated InSb quantum wire without the need of using magnetic fields, static or oscillating. 
We presented two nanodevices: one capable to rotate the electron spin around one axis, and the other nanodevice is able to rotate the electron spin around two orthogonal axis which means that it allows to realize any single spin rotation. Both nanodevices are all electrically controlled by oscillating voltages applied to the electrodes and act in the ultrafast manner (subnanoseconds). 
The oscillations of the Rashba spin-orbit coupling shifted in phase by $\pi/2$ between the electron position oscillations leads to the spin rotations. The results are supported by precise numerical simulations of the 3D Poisson-Schr\"{o}dinger type which allows to include various electrostatic effects and nonuniform 3D RSOI. Electrostatic study showed that the presence of the induced charge on the interface between materials with different electric permittivities can significantly influence shape of the confinement potential. The simulations were performed for the structures which are within the reach of the current experiments.




\begin{acknowledgments}
This work has been supported by National Science Centre, Poland, under Grant No. UMO-2014/13/B/ST3/04526. P.~S. acknowledge support from SCIEX. 

\end{acknowledgments}

\appendix*

\section{Influence of the various components on the shape of the confinement potential }
The voltages applied to the electrodes have important influence on the confinement potential shape, also in the $z$ direction.
Spatially nonuniform  electric permittivity $\epsilon(\mathbf{r})$
which leads to appearance of induced charges on the interfaces between different materials is also very important to determine the proper shape of the confinement potential. However, it turns out that contribution of the electron gas charge $\rho_\mathrm{eg}$ has a much smaller influence on the confinement shape.

In order to tell which of these two charge sources (electron gas and the charge induced on the different permittivities border) have a major impact on the confinement shape, 
we have repeated the calculations presented above with modified versions of the considered Poisson equation---Eq.~(\ref{poiss1}).
In the first case (A), in our calculations we do not take into account the electron gas and charge induced on the interfaces between different materials from the system by considering the following equation:
\begin{equation}\tag{A}
\epsilon_0\epsilon(\mathbf{r})\boldsymbol{\nabla}^2 \Phi(\mathbf{r})=-\rho_\mathrm{e}(\mathbf{r}).
\end{equation}
In the second case (B), we include only influence of the charge induced on the interfaces:
\begin{equation}\tag{B}
\boldsymbol\nabla \cdot \left(\epsilon_0\epsilon(\mathbf{r})\boldsymbol\nabla \Phi(\mathbf{r})\right)=-\rho_\mathrm{e}(\mathbf{r}),
\end{equation}
in third case (C), only the electron gas is included:
\begin{equation}\tag{C}
\epsilon_0\epsilon(\mathbf{r})\boldsymbol\nabla^2 \Phi(\mathbf{r})=-\left(\rho_\mathrm{e}(\mathbf{r})+\rho_\mathrm{eg}(\mathbf{r})\right).
\end{equation}
Everything is compared with solutions for the original version of the Poisson equation~(D):
\begin{equation}\tag{D}
\boldsymbol\nabla \cdot \left(\epsilon_0\epsilon(\mathbf{r})\boldsymbol\nabla \Phi(\mathbf{r})\right)=-\left(\rho_\mathrm{e}(\mathbf{r})+\rho_\mathrm{eg}(\mathbf{r})\right).
\end{equation}

\begin{figure}[]
\includegraphics[angle=-90,width=9.0cm]{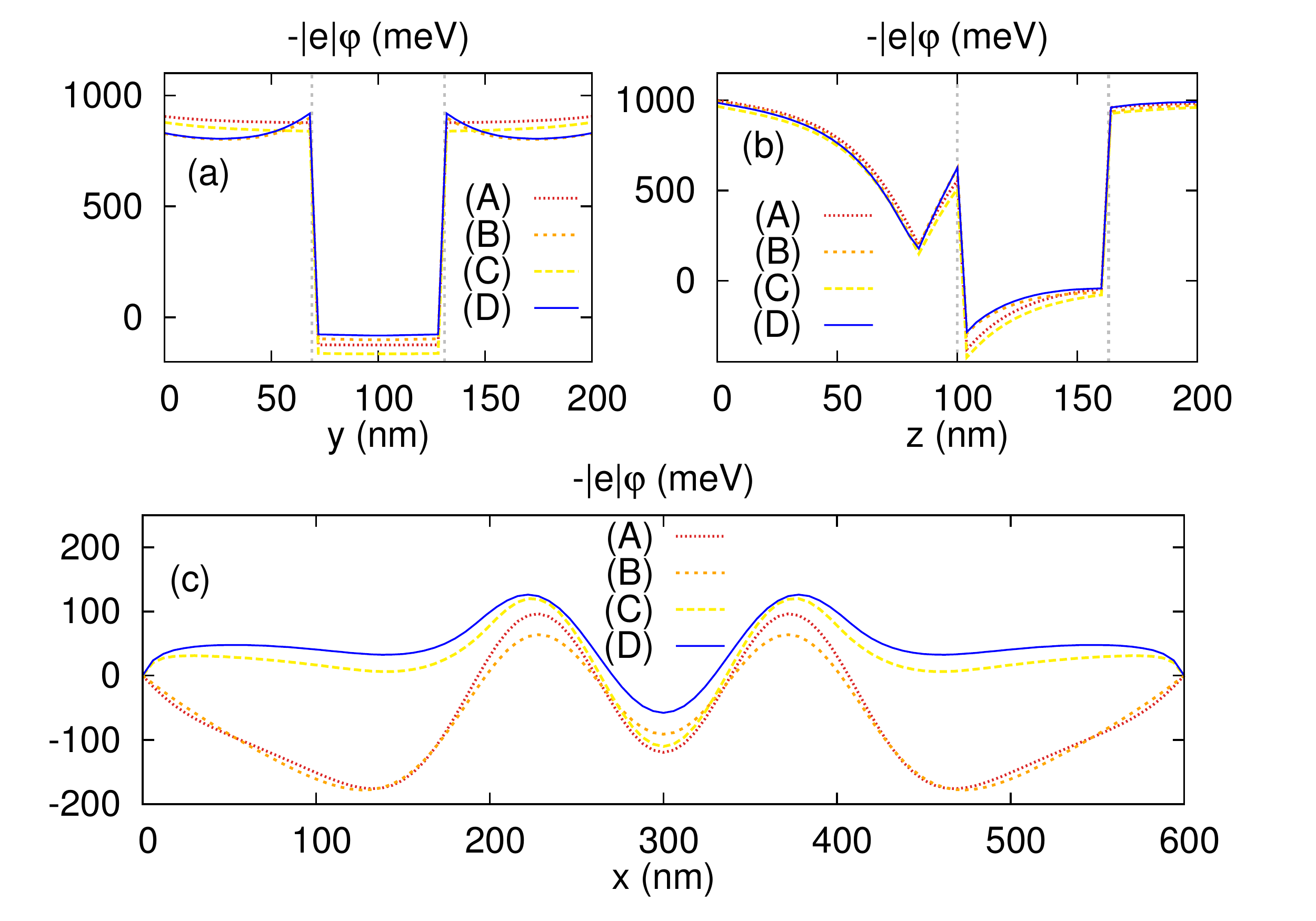}
\caption{\label{fig:19} (color online). Same as in Fig.~\ref{fig:4}, but here confinement potential calculated with (A-D) Poisson's equations forms denoted by red (dotted line), 
orange (long dashes), yellow (short dashes) and blue (solid line) colors respectively.}
\end{figure}
In all the cases (A-D) we assume presence of the electron in the quantum dot.
The results---confinement potential (cross-section along $x$, $y$, and $z$ axis) for different forms of Poisson's equation---cases (A-D) are depicted in Fig.~\ref{fig:19} (in similar manner as in Fig.~\ref{fig:4}).
In the lateral confinement cases (subfigures (a) and (b)), we see that taking into account the electron gas does not change confinement significantly (case (C) vs. (A)), 
while the induced charge on interfaces affects substantially the lateral  confinement (case (B) vs. (A)) leading to a confinement potential shape similar to the original (D). 
Confinement along the wire (subfigure (c)) in the dot region (localized in the center of the wire ) in case (B) has only small offset relative to the original (D).

Simple image charge method for half spaces with permittivities $\epsilon$ and $\epsilon'$ states that:
if on the $\epsilon'$ side a charge $q$ emerges it will feel the mirror charge with a value of $-q(\epsilon-\epsilon')/(\epsilon+\epsilon')$. 
Thus, if $\epsilon<\epsilon'$ then the image charge will have the same sign as $q$. 
The charge $q$ seen from the $\epsilon$ side has increased (for $\epsilon<\epsilon'$) value $2q\epsilon'/(\epsilon+\epsilon')$.
We have such case here: the electron and gas charge is located in the wire region with higher permitivity than the permitivity of the surrounding area.
This causes potential level bending up on both wire border sides, which is best seen in subfigure (a).
The induced charge significantly reduces the confinement potential slope in $z$ direction---subfigure (b), 
which changes the Rashba coupling: here from $53$~meV$\,$nm for the case (C) to $37$~meV$\,$nm for the case (D).


\bibliography{drutbezpola} 

\end{document}